\documentclass[12pt]{article}
\usepackage{amsmath}

\makeatletter

\@addtoreset{equation}{section}

\def\draftdate{\relax}
\def\mda{\relax}
\def\mua{\relax}
\def\mla{\relax}
\def\draft{
\def\thtystars{******************************}
\def\sixtystars{\thtystars\thtystars}
\typeout{}
\typeout{\sixtystars**}
\typeout{* Draft mode!
         For final version remove \protect\draft\space in source file *}
\typeout{\sixtystars**}
\typeout{}
\def\draftdate{\today}
\def\mua{\marginpar[\boldmath\hfil$\uparrow$]%
                   {\boldmath$\uparrow$\hfil}%
                    \typeout{marginpar: $\uparrow$}\ignorespaces}
\def\mda{\marginpar[\boldmath\hfil$\downarrow$]%
                   {\boldmath$\downarrow$\hfil}%
                    \typeout{marginpar: $\downarrow$}\ignorespaces}
\def\mla{\marginpar[\boldmath\hfil$\rightarrow$]%
                   {\boldmath$\leftarrow $\hfil}%
                    \typeout{marginpar: $\leftrightarrow$}\ignorespaces}
\overfullrule 5pt
\oddsidemargin -15mm
\marginparwidth 29mm
}

\def\stars{\strut\leaders\hbox{*}\hfill\strut}
\def\starline{\hfil\strut\hfil\hbox to \textwidth {\stars}\hfil}

\newcommand\Ref[1]     {Ref.\,\cite{#1}}
\newcommand\Refs[1]    {Refs.\,\cite{#1}}
\newcommand\eqn[1]     {Eq.\,(\ref{#1})}
\newcommand\eqns[2]    {Eqs.\,(\ref{#1}) and~(\ref{#2})}
\newcommand\eqnss[2]   {Eqs.\,(\ref{#1})--(\ref{#2})}

\newcommand\app[1]     {Appendix~\ref{#1}}
\newcommand\tab[1]     {Table~\ref{#1}}
\newcommand\nn         {\nonumber}

\def\beq{\begin{equation}}
\def\eeq{\end{equation}}
\def\beeq{\begin{eqnarray}}
\def\eeeq{\end{eqnarray}}
\newcommand\bom[1]     {{\mbox{\boldmath $#1$}}}
\def\aand{\!\!\!\!\!\!\!\!&&}

\newcommand\as         {\ensuremath{\alpha_{\mathrm{s}}}}
\newcommand\asb        {\ensuremath{\alpha_{\mathrm{s}}^{\mathrm u}}}
\newcommand\gs         {\ensuremath{g_{\mathrm{s}}}}
\newcommand\aeps{\frac{\as}{2\pi}\,S_\eps\left(\frac{\mu^2}{Q^2}\right)^{\eps}}

\newcommand{\CF}       {C_{\mathrm{F}}}
\newcommand{\CA}       {C_{\mathrm{A}}}
\newcommand{\TR}       {T_{\mathrm{R}}}
\newcommand{\Nc}       {N_{\mathrm{c}}}
\newcommand{\Nf}       {n_{\mathrm{f}}}

\newcommand{\bT}       {\bom{T}}

\newcommand\qb         {{\bar q}}
\newcommand\msbar      {\ensuremath{{\overline {\rm MS}}}}
\newcommand\muR[1]     {\ensuremath{\mu_R^{#1}}}
\newcommand\e          {{\mathrm e}}

\renewcommand\O        {{\mathrm O}}
\newcommand\Oe[1]      {\ensuremath{\mathrm O(\eps^{#1})}}
\newcommand{\eps}      {\varepsilon}        
\newcommand{\finite}   {{\cal F}\!in}        

\newcommand\Li         {\mathop{\mathrm{Li}}\nolimits}
\newcommand\Real       {\mathop{\mathrm{Re}}\nolimits}

\newcommand\ldot       {\!\cdot\!}
\newcommand\smfrac[2]  {{\textstyle\frac{#1}{#2}}}
\newcommand\hf         {\smfrac{1}{2}}

\newcommand{\PS}[1]    {\rd\phi_{#1}}
\newcommand{\rd}{{\mathrm{d}}}
\newcommand\tsig[1]    {\sigma^{\mathrm{#1}}}
\newcommand\dsig[1]    {\rd\sigma^{{\rm #1}}}
\newcommand\dsiga[2]   {\rd\sigma^{{\rm #1,A}_{\scriptscriptstyle #2}}}
\newcommand{\Jac}[1]   {{\cal J}_m^{#1}}

\newcommand\la         {\langle}
\newcommand\ra         {\rangle}

\newcommand{\cA}       {{\cal A}}
\newcommand{\cM}       {{\cal M}}
\newcommand\SME[3]     {|{\cal M}_{#1}^{(#2)}{(#3)}|^2}
\newcommand\M[2]       {\ensuremath{|{\cal{M}}_{#1}^{#2}|^2}}
\newcommand\bra[3]     {\la {\cal M}_{#1}^{#2}#3|}
\newcommand\ket[3]     {|{\cal M}_{#1}^{#2}#3\ra}

\newcommand{\mom}[1]   {\{p\}^{#1}}
\newcommand{\momt}[1]   {\{\ti{p}\}^{#1}}

\newcommand{\bA}[1]    {\bom{\mathrm A}_{#1}}
\newcommand{\bC}[1]    {\bom{\mathrm C}_{#1}}

\newcommand{\bS}[1]    {\bom{\mathrm S}_{#1}}

\newcommand{\bCS}[2]   {\bom{\mathrm C}_{#1}\bom{\mathrm S}_{#2}}

\newcommand{\bSCS}[1]  {\bom{\mathrm C}\kern-2pt\bom{\mathrm S}_{#1}}

\def\hP{\hat{P}}
\newcommand{\calS}     {{\cal S}}
\newcommand{\bcA}[2]   {{\bom{\cal A}}_{#1}^{#2}}
\newcommand{\cC}[2]    {{\cal C}_{#1}^{#2}}
\newcommand{\cS}[2]    {{\cal S}_{#1}^{#2}}
\newcommand{\cCS}[3]   {{\cal C}_{#1}^{~}{\cal S}_{#2}^{#3}}

\newcommand{\cSCS}[1]  {{\cal C}\kern-2pt{\cal S}_{#1}^{~}}

\newcommand{\IcC}[2]   {{\mathrm C}_{#1}^{#2}}
\newcommand{\IcS}[2]   {{\mathrm S}_{#1}^{#2}}
\newcommand{\IcCS}[1]  {\mathrm{C\!S}^{#1}}

\newcommand{\bI}       {\bom{I}}

\newcommand{\ti}[1]    {\tilde{#1}}
\newcommand{\wti}[1]   {\widetilde{\,#1\,}}

\newcommand\tzz[2]     {z_{#1,#2}}

\newcommand\kT[1]      {k_{\perp,#1}}
\newcommand\kTt[1]     {k_{\perp,#1}}

\begin{document}


\begin{titlepage}
\renewcommand{\thefootnote}{\fnsymbol{footnote}}
\begin{flushright}
hep-ph/0609041 
     \end{flushright}
\par \vspace{5mm}
\begin{center}
{\Large \bf A new subtraction scheme for computing QCD jet cross sections
at next-to-leading order accuracy} \\[.5cm] 
{\large \em Dedicated to Professor Istv\'an Lovas on his 75th birthday}
\end{center}

\par \vspace{2mm}
\begin{center}
{\bf G\'abor Somogyi} and {\bf Zolt\'an Tr\'ocs\'anyi}\\[.5em]
{University of Debrecen and \\Institute of Nuclear Research of
the Hungarian Academy of Sciences\\ H-4001 Debrecen, PO Box 51, Hungary}
\end{center}

\par \vspace{2mm}
\begin{center} {\large \bf Abstract} \end{center}
\begin{quote}
\pretolerance 10000
We present a new subtraction scheme for computing jet cross sections
in electron-positron annihilation at next-to-leading order accuracy in
perturbative QCD. The new scheme is motivated by problems emerging in
extending the subtraction scheme to the next-to-next-to-leading order.
The new scheme is tested by comparing predictions for three-jet
event-shape distributions to those obtained by the standard program
{\sc event}.
\end{quote}

\vspace*{\fill}
\begin{flushleft}
September 2006
\end{flushleft}
\end{titlepage}
\clearpage

\tableofcontents

\renewcommand{\thefootnote}{\fnsymbol{footnote}}


%
%

\section{Introduction}
\label{sec:intro}

High-energy physics will enter a new era of discovery with the start
of LHC operations in 2007.  The LHC is a proton-proton collider that
will  function at the highest energy ever attained in the laboratory, 
and will probe  a new realm of high-energy physics.  The use of a
high-energy hadron collider as a research tool makes substantial
demands upon the theoretical understanding and predictive power of 
QCD, the theory of the strong interactions within the Standard Model.  
  
At high $Q^2$ any production rate can be expressed as a series
expansion  in $\alpha_S$. Because QCD is asymptotically free, the
simplest approximation  is to evaluate any series expansion to leading
order in $\alpha_S$.  However, for most processes a leading-order
evaluation yields unreliable   predictions. The next simplest
approximation is the inclusion of radiative corrections at the
next-to-leading order (NLO) accuracy, which  usually warrants a
satisfying assessment of the production rates. In the previous decade,
a lot of effort was devoted  to devise process-independent methods and
compute rates to NLO accuracy and the problem is known to be solved
\cite{Giele:1991vf, Giele:1993dj,Frixione:1995ms,Nagy:1996bz,%
Frixione:1997np,Catani:1996vz}.  In particular, the dipole subtraction
scheme \cite{Catani:1996vz} provides a simple and fully universal way
of computing the radiative corrections and has been implemented in two
widely used programs the {\sc mcfm}
\cite{Campbell:2000bg,Campbell:2002tg,Campbell:2004ch} and the {\sc
nlojet++} \cite{Nagy:2003tz,Nagy:2001xb,Nagy:1998bb} codes.

In some cases, typically and most importantly when the NLO corrections
are large, the corrections at next-to-next-to-leading order (NNLO)
accuracy are necessary in order to give a reliable prediction of the
production rates.  Recently, a lot of effort has been devoted to
the extension of the subtraction schemes used at NLO to NNLO. It was
found however, that the universal NLO schemes cannot be extended to
NNLO \cite{Somogyi:2006_2,Somogyi:2006_3}, which motivates the new
method presented in this paper.

\section{Subtraction scheme at NLO}
\label{sec:sub_NLO}

The jet cross sections in perturbative QCD are represented by an
expansion in the strong coupling $\as$. At NLO accuracy we keep the
two lowest-order terms,
\beq
\sigma_{\rm NLO} = \tsig{LO} + \tsig{NLO}\,.
\eeq
Assuming an $m$-jet quantity, the leading-order contribution is the
integral of the fully differential Born cross section $\dsig{B}_m$ of
$m$ final-state partons over the available $m$-parton phase space
defined by the jet function $J_m$,
\beq
\tsig{LO} = \int_m\!\dsig{B}_m J_m\:.
\eeq
The NLO contribution is a sum of two terms, the real and virtual
corrections,
\beq
\tsig{NLO} =
\int_{m+1}\!\dsig{R}_{m+1} J_{m+1} + \int_m\!\dsig{V}_m J_m\:.
\label{eq:sigmaNLO}
\eeq
Here the notation for the integrals indicates that the real correction
involves $m+1$ final-state partons, one of those being unresolved, while
the virtual correction has $m$-parton kinematics, and the phase
spaces are restricted by the corresponding jet functions $J_n$ that
define the physical quantity.

In $d=4$ dimensions the two contributions in  \eqn{eq:sigmaNLO}
are separately divergent, but their sum is finite for infrared-safe
observables order by order in the expansion in $\as$. The requirement
of infrared-safety puts constraints on the analytic behaviour of the
jet functions that were spelled out explicitly in \Ref{Catani:1996vz}.

The traditional approach to finding the finite corrections at NLO
accuracy is to first continue all integrations to $d = 4 -2\eps$ ($\eps
\ne 0$) dimensions, then regularise the real radiation contribution by
subtracting a suitably defined approximate cross section
$\dsig{R,A}_{m+1}$ such that (i) $\dsig{R,A}_{m+1}$ matches the
point-wise singular behaviour of $\dsig{R}$ in the one-parton IR
regions of the phase space in any dimensions (ii) and it can be
integrated over the one-parton phase space of the unresolved parton
independently of the jet function, resulting in a Laurent expansion in
$\eps$. After performing this integration, the approximate cross
section can be combined with the virtual correction $\dsig{V}$ before
integration. We then write 
\beq
\tsig{NLO} =
  \int_{m+1}\!
  \left[\dsig{R}_{m+1} J_{m+1} - \dsig{R,A}_{m+1} J_m\right]_{\eps = 0}
+ \int_m\!
  \left[\dsig{V}_m + \int_1\!  \dsig{R,A}_{m+1}\right]_{\eps = 0} J_m\:.
\label{eq:sigmaNLO2}
\eeq
(Note that $\dsiga{R}{}_{m+1}$ is multiplied by $J_m$, therefore, after
integration over the phase space of the unresolved parton, it can be
combined with $\dsig{V}_{m}$.) Since the first integral on the right
hand side of \eqn{eq:sigmaNLO2} is finite in $d=4$ dimensions
by construction, it follows from the Kinoshita-Lee-Nauenberg theorem
that the combination of terms in the $m$-parton integral is finite as
well, provided the jet function defines an infrared-safe observable.

The final result is that one is able to rewrite the two NLO
contributions in \eqn{eq:sigmaNLO} as a sum of two finite integrals,
\beq
\tsig{NLO} =
\int_{m+1}\!\dsig{NLO}_{m+1} + \int_m\!\dsig{NLO}_m\:,
\label{eq:sigmaNLOfin}
\eeq
that are integrable in four dimensions using standard numerical
techniques.

\section{Notation}
\label{sec:notation}

\subsection{Matrix elements}
\label{sec:ME}

We consider processes with coloured particles (partons) in the final
states, while the initial-state particles are colourless (typically
electron-positron annihilation into hadrons).  Any number of additional
non-coloured final-state particles are allowed, too, but they will be
suppressed in the notation.  Resolved partons in the final state are
labelled by $i,k,l,\dots$, the unresolved one is denoted by $r$.

We adopt the colour- and spin-state notation of \Ref{Catani:1996vz}. In
this notation the amplitude for a scattering process involving the
final-state momenta $\mom{}$, $\ket{m}{}{(\mom{})}$, is an abstract
vector in colour and spin space, and its normalization is fixed such
that the squared amplitude summed over colours and spins is
\beq
\label{eq:M2}
|\cM_m|^2 = \bra{m}{}{}\ket{m}{}{}\:.
\eeq
This matrix element has the following formal loop-expansion:
\beq
\ket{}{}{} = \ket{}{(0)}{} + \ket{}{(1)}{} + \dots\,,
\label{FormalLoopExpansion}
\eeq
where $\ket{}{(0)}{}$ denotes the tree-level contribution,
$\ket{}{(1)}{}$ is the one-loop contribution and the dots stand for
higher-loop contributions, which are not used in this paper. 

Colour interactions at the QCD vertices are represented by associating
colour charges $\bT_i$ with the emission of a gluon from each
parton $i$.  In the colour-state notation, each vector $\ket{}{}{}$ is
a colour-singlet state, so colour conservation is simply
\beq
\biggl(\sum_j \bT_j \biggr) \,\ket{}{}{} = 0\,,
\eeq
where the sum over $j$ extends over all the external partons of the
state vector $\ket{}{}{}$, and the equation is valid order by order in
the loop expansion of \eqn{FormalLoopExpansion}. 

Using the colour-state notation, we can write the two-parton
colour-correlated squared tree amplitudes as
\beeq
|\cM^{(0)}_{(i,k)}(\{p\})|^2 \aand \equiv
\bra{}{(0)}{(\{p\})} \,\bT_i \ldot \bT_k \, \ket{}{(0)}{(\{p\})}
\:.
\label{eq:colam2}
\eeeq
The colour-charge algebra for the product 
$(\bT_i)^n (\bT_k)^n \equiv \bT_i \ldot \bT_k$ is:
\beq
\bT_i \ldot \bT_k =\bT_k \ldot \bT_i \quad  {\rm if}
\quad i \neq k; \qquad \bT_i^2= C_i\:.
\label{eq:colalg}
\eeq
Here $C_i$ is the quadratic Casimir operator in the representation of
particle $i$ and we have $\CF= \TR(\Nc^2-1)/\Nc= (\Nc^2-1)/(2\Nc)$ in
the fundamental and $\CA=2\,\TR \Nc=\Nc$ in the adjoint representation,
i.e.~we are using the customary normalization $\TR=1/2$.

\subsection{Dimensional regularization, one-loop amplitudes
and renormalization}
\label{sec:dimreg}

We employ conventional dimensional regularization (CDR) in $d=4-2\eps$
space-time dimensions to regulate both the IR and UV divergences,
when quarks (spin-$\hf$ Dirac fermions) possess 2 spin polarizations,
gluons have $d-2$ helicity states and all particle momenta are taken as
$d$-dimensional.

Turning to the renormalization of the amplitudes, let the perturbative
expansion of the unrenormalised scattering amplitude $|\cA_m\ra$ in terms
of the bare coupling $\gs \equiv \sqrt{4\pi \asb}$ be
\beq
|\cA_m\ra = \left(\frac{\asb\mu^{2\eps}}{4\pi}\right)^{q/2}
\left[
  |\cA_m^{(0)}\ra
+ \left(\frac{\asb\mu^{2\eps}}{4\pi}\right) |\cA_m^{(1)}\ra
+ \O\left( (\alpha_{\mathrm{s}}^{\mathrm u})^2\right)
\right]\,,
\label{cA}
\eeq
where $q$ is a non-negative integer and $\mu$ is the
dimensional-regularization scale. The renormalized amplitudes
$|\cM_m\ra$ are obtained from the unrenormalized ones by expressing the
bare coupling in terms of the running coupling $\as(\muR2)$ evaluated
at the arbitrary renormalization scale $\muR2$ as
\beq
\asb \mu^{2\eps} = \as(\muR2)\,\muR{2\eps}\,S_\eps^{-1}
\left[1 - \left(\frac{\as(\muR2)}{4\pi}\right)\frac{\beta_0}{\eps}
+ \O(\as^2(\muR2))\right]\,,
\label{couplingrenormalization}
\eeq
where $\beta_0$ is the first coefficient of the QCD
$\beta$ function for $\Nf$ number of light quark flavours,
\beq
\beta_0 = \frac{11}3\CA - \frac43\TR\Nf
\,.
\eeq
In \eqn{couplingrenormalization}, $S_\eps$ is the phase space factor
due to the integral over the $(d-3)$-dimensional solid angle, which is
included in the definition of the running coupling in the \msbar\
renormalization scheme,\footnote{The \msbar\ renormalization scheme as
often employed in the literature uses $S_\eps=(4\pi)^\eps \e^{-\eps\gamma_E}$.
It is not difficult to check that the two definitions lead to the same
expressions in a computation at the NLO accuracy.}
\beq
S_\eps = \int\!\frac{\rd^{(d-3)}\Omega}{(2\pi)^{d-3}} =
\frac{(4\pi)^\eps}{\Gamma(1-\eps)}\,.
\eeq
We always consider the running coupling in the \msbar\ scheme defined
with the inclusion of this phase space factor.

The relations between the renormalized amplitudes of
\eqn{FormalLoopExpansion} and the unrenormalized ones are given as
follows:
\beeq
&&
\ket{m}{(0)}{} =
\left(\frac{\as(\muR2)\,\muR{2\eps}}{4\pi} S_\eps^{-1}\right)^{q/2}
|\cA_m^{(0)}\ra
\label{renA0}
\\ &&
\ket{m}{(1)}{} =
\left(\frac{\as(\muR2)\,\muR{2\eps}}{4\pi} S_\eps^{-1}\right)^{q/2}
\frac{\as(\muR2)}{4\pi} S_\eps^{-1}
\left(\muR{2\eps}\,|\cA_m^{(1)}\ra
 - \frac{q}{2}\frac{\beta_0}{\eps} S_\eps\,|\cA_m^{(0)}\ra\right)
\,.
\label{renA1}
\eeeq

After UV renormalization, the dependence on $\mu$ turns into a
dependence on $\muR{}$, so the physical cross sections depend only on
the renormalization scale $\muR{}$.  To avoid a cumbersome notation, we
therefore set $\muR{}=\mu$ in the rest of the paper. Furthermore, after
the IR poles are canceled in an NLO computation we may set $\eps = 0$,
therefore, the $\muR{2\eps}$ and $S_\eps^{-1}$ factors that accompany
the running coupling in the renormalized amplitude do not give any
contribution, so we may perform the 
\beq
\left(\frac{\as(\muR2)\,\muR{2\eps}}{4\pi} S_\eps^{-1}\right)^{q/2}
\left(\frac{\as(\muR2)}{4\pi} S_\eps^{-1}\right)^i \to
\left(\frac{\as(\muR2)}{4\pi}\right)^{q/2+i}
\eeq
substitution in \eqnss{renA0}{renA1}.

\subsection{Cross sections}
\label{sec:xsecs}

In our notation the real cross section $\dsig{R}_{m+1}$ is
given by
\beq
\dsig{R}_{m+1}= {\cal N}\sum_{\{m+1\}}\PS{m+1}{}(\mom{})
\,\frac{1}{S_{\{m+1\}}}
\,\la {\cal M}_{m+1}^{(0)}(\mom{})|{\cal M}_{m+1}^{(0)}(\mom{}) \ra\,,
\label{eq:dsigR}
\eeq
where ${\cal N}$ includes all QCD-independent factors, $\sum_{\{m+1\}}$
denotes a summation over all subprocesses, $S_{\{m+1\}}$ is the Bose
symmetry factor for identical particles in the final state and
$\PS{m+1}{(\mom{})}$ is the $d$-dimensional phase space for
$m+1$ outgoing particles with momenta
$\mom{} \equiv \{p_1,\dots,p_{m+1}\}$ and total momentum $Q$,
\beq
\label{eq:psf}
\rd\phi_{m+1}(p_1,\dots,p_{m+1};Q) = 
\left[ \,\prod_{i=1}^{m+1} \frac{\rd^{d}p_i}{(2\pi)^{d-1}}
\,\delta_+(p_i^2) \right] 
(2\pi)^d \delta^{(d)}(p_1+\dots+p_{m+1}-Q)\,.
\eeq
The virtual contribution $\dsig{V}_{m}$ is
\beq
\dsig{V}_{m} = 
{\cal N}\sum_{\{m\}}\PS{m}{}
\,\frac{1}{S_{\{m\}}}
\:2 \Real \la {\cal M}_{m}^{(0)}|{\cal M}_{m}^{(1)} \ra\,.
\label{eq:dsigV}
\eeq
In the rest of the paper, we define explicitly the approximate cross
section $\dsig{R,A}_{m+1}$ and compute its integral
$\int_1 \dsiga{R}1_{m+1}$.

\section{The approximate cross section}
\label{sec:int_RRA1}

The construction of the suitable approximate cross section
$\dsig{R,A}_{m+1}$ is made possible by the universal soft and collinear
factorization properties of QCD matrix elements
\cite{Altarelli:1977zs,Mangano:1990by}.  In \Ref{Somogyi:2005xz} we
introduced symbolic operators $\bC{ir}$, $\bS{r}$ that perform the
action of taking the collinear limit ($p^\mu_i || p^\mu_r$), or soft limit%
\footnote{For the precise definition of the collinear and soft limits,
refer to \Refs{Catani:1996vz,Somogyi:2005xz}.}
($p^\mu_r \to 0$) of the squared matrix elements, respectively, keeping
the leading singular term. Using this notation, we defined the formal
expression $\bA{}|\cM_{m+1}^{(0)}|^2$, that matches the singular
behaviour of the squared matrix element in all the singly-unresolved
regions of the phase space,
\beeq
&&
\bA{}|\cM_{m+1}^{(0)}|^2 =
\sum_{r} \left[\sum_{i\ne r} \frac12 \bC{ir}
+ \left(\bS{r} - \sum_{i\ne r} \bCS{ir}{r}\right) \right]
|\cM_{m+1}^{(0)}{(p_i,p_r,\dots)}|^2
\,.
\label{eq:bA1}
\eeeq
This expression cannot yet serve as a subtraction term because it is
defined precisely only in the strict collinear and/or soft limits. It
has to be extended over the whole phase space, which requires an exact
factorization of the $m+1$ parton phase space into an $m$ parton phase
space times the phase space measure of the unresolved parton,
\beq
\PS{m+1}{(\mom{})} = \PS{m}{(\momt{})}\,[\rd p_1]\,.
\label{eq:PS_fact}
\eeq
With this phase-space factorization we define the approximate cross
section as
\beq
\dsig{R,A}_{m+1} = \PS{m}{}\:[\rd p_{1}]\:\bcA{}{} \M{m+1}{(0)}\,,
\label{eq:dsigRA1}
\eeq
where $\bcA{}{} \M{m+1}{(0)}$ has the same structure as \eqn{eq:bA1},
\beq
\bcA{}{}\SME{m+1}{0}{\mom{}} =
\sum_{r} \left[\sum_{i\ne r} \frac{1}{2} \cC{ir}{}(\mom{})
+ \left(\cS{r}{}(\mom{}) - \sum_{i\ne r} \cCS{ir}{r}{}(\mom{})\right) 
\right]\,.
\label{eq:bcA1}
\eeq
We now define all terms on the right hand side of \eqn{eq:bcA1}
precisely.

The collinear counterterm $\cC{ir}{}(\mom{})$ reads
\beq
\cC{ir}{}(\mom{}) = 
8\pi\as\mu^{2\eps}\frac{1}{s_{ir}}
\bra{m}{}{(\momt{(ir)})}
\hP_{f_i f_r}^{}(\tzz{i}{r},\tzz{r}{i},\kTt{i,r};\eps)
\ket{m}{}{(\momt{(ir)})}\,,
\label{eq:Cir00}
\eeq
where the $\hP_{f_i f_r}^{}(\tzz{i}{r},\tzz{r}{i},\kTt{i,r};\eps)$ kernels 
are the $d$-dimensional Altarelli-Parisi splitting functions as given
in \Ref{Somogyi:2005xz}.%
\footnote{Note in particular that the ordering of the flavour indices and
arguments of the Altarelli-Parisi kernels has no meaning in the notation
of \Ref{Somogyi:2005xz}.}
The momentum fractions $\tzz{i}{r}$ and
$\tzz{r}{i}$ are 
\beq
\tzz{i}{r} = \frac{y_{iQ}}{y_{(ir)Q}}
\qquad\mbox{and}\qquad
\tzz{r}{i} = \frac{y_{rQ}}{y_{(ir)Q}}\,,
\label{eq:zt2}
\eeq
while the transverse momentum $\kTt{i,r}$ is
\beq
\kTt{i,r}^{\mu} = 
\zeta_{i,r} p_r^{\mu} - \zeta_{r,i} p_i^{\mu} + \zeta_{ir} \ti{p}_{ir}^{\mu}
\,,\qquad
\zeta_{i,r} = \tzz{i}{r}-\frac{y_{ir}}{\alpha_{ir}y_{(ir)Q}}
\,,\quad
\zeta_{r,i} =  \tzz{r}{i}-\frac{y_{ir}}{\alpha_{ir}y_{(ir)Q}}
\,.
\label{eq:kTtir}
\eeq
We used the abbreviations $y_{ir}= s_{ir}/Q^2 \equiv 2p_i\cdot p_r/Q^2$,
$y_{(ir)Q} = y_{iQ} + y_{rQ}$ with $y_{iQ}=2p_i\cdot Q/Q^2$,
$y_{rQ}=2p_r\cdot Q/Q^2$ and $Q^\mu$ is the total four-momentum of the
incoming electron and positron, while  $\ti{p}_{ir}^{\mu}$ and
$\alpha_{ir}$ are defined below in \eqns{eq:PS_Cir}{eq:alphair}
respectively.  This choice for the transverse momentum is exactly
perpendicular to the parent momentum $\ti{p}_{ir}^{\mu}$ and ensures
that in the collinear limit $p_i^\mu || p_r^\mu$, the square of
$\kTt{i,r}^{\mu}$ behaves as
\beq
\kTt{i,r}^2 \simeq - s_{ir}\tzz{r}{i} \tzz{i}{r}
\,,
\label{eq:kTir2}
\eeq
as required (independently of $\zeta_{ir}$). Choosing 
\beq
\zeta_{ir} =
\frac{y_{ir}}{\alpha_{ir} y_{\wti{ir}Q}}(\tzz{r}{i}-\tzz{i}{r})\,,
\label{eq:zetair}
\eeq
$\kTt{i,r}^\mu \to \kT{i}^\mu$ in the collinear limit as can be shown
by substituting the Sudakov parametrization of the momenta into
\eqn{eq:kTtir} (with properly chosen gauge vector). Note however, that
in a NLO computation, fulfilling \eqn{eq:kTir2} is sufficient to ensure
the correct collinear behaviour of the subtraction term and the
longitudinal component that is proportional to $\zeta_{ir}$ does not
contribute due to gauge invariance of the matrix elements, so we may
choose $\zeta_{ir} = 0$. The $m$ momenta $\momt{(ir)} \equiv
\{\ti{p}_1,\ldots,\ti{p}_{ir},\ldots,\ti{p}_{m+1}\}$ entering the
matrix elements on the right hand side of \eqn{eq:Cir00} are
\beq
\ti{p}_{ir}^{\mu} =
\frac{1}{1-\alpha_{ir}}(p_i^{\mu} + p_r^{\mu} - \alpha_{ir} Q^{\mu})\,,
\qquad
\ti{p}_n^{\mu} = \frac{1}{1-\alpha_{ir}} p_n^{\mu}\,,
\qquad n\ne i,r\,,
\label{eq:PS_Cir}
\eeq
where
\beq
\alpha_{ir} =
\frac12\Big[y_{(ir)Q}-\sqrt{y_{(ir)Q}^2 - 4y_{ir}}\;\Big]\,.
\label{eq:alphair}
\eeq

The soft and soft-collinear counterterms $\cS{r}{}(\mom{})$ and 
$\cCS{ir}{r}{}(\mom{})$ are
\beeq
\cS{r_g}{}(\mom{}) \aand= 
-8\pi\as\mu^{2\eps}\sum_{i}\sum_{k\ne i} \frac12 \calS_{ik}(r)
\SME{m,(i,k)}{0}{\momt{(r)}}\,,
\label{eq:Sr00}
\\
\cCS{ir_g}{r_g}{}(\mom{}) \aand= 
8\pi\as\mu^{2\eps} \frac{1}{s_{ir}}\frac{2\tzz{i}{r}}{\tzz{r}{i}}\,\bT_i^2\,
\SME{m}{0}{\momt{(r)}}\,.
\label{eq:CirSr00}
\eeeq
If $r$ is a quark or antiquark, $\cS{r}{}(\mom{})$ and $\cCS{ir}{r}{}(\mom{})$
are both zero. The eikonal factor in \eqn{eq:Sr00} is
\beq
\calS_{ik}(r) = \frac{2 s_{ik}}{s_{ir} s_{rk}}\,,
\label{eq:Sikr}
\eeq
and the momentum fractions entering \eqn{eq:CirSr00} are given in \eqn{eq:zt2}.
The $m$ momenta
$\momt{(r)} \equiv \{\ti{p}_1,\ldots,\ti{p}_{m+1}\}$ ($p_r$ is absent)
entering the matrix elements on the right hand sides of 
\eqns{eq:Sr00}{eq:CirSr00} read
\beq
\ti{p}_n^{\mu} =
\Lambda^{\mu}_{\nu}[Q,(Q-p_r)/\lambda_r] (p_n^{\nu}/\lambda_r)\,,
\qquad n\ne r\,,
\label{eq:PS_Sr}
\eeq
where
\beq
\lambda_r = \sqrt{1-y_{rQ}}\,,
\label{eq:lambdar}
\eeq
and
\beq
\Lambda^{\mu}_{\nu}[K,\wti{K}] = g^{\mu}_{\nu}
- \frac{2(K+\wti{K})^{\mu}(K+\wti{K})_{\nu}}{(K+\wti{K})^{2}} 
+ \frac{2K^{\mu}\wti{K}_{\nu}}{K^2}\,.
\label{eq:LambdaKKt}
\eeq
The matrix $\Lambda^{\mu}_{\nu}[K,\wti{K}]$ generates a (proper) Lorentz
transformation, provided $K^2 = \wti{K}^2 \ne 0$.

The momentum mappings introduced in \eqns{eq:PS_Cir}{eq:PS_Sr} both
lead to exact phase space factorization in the form of
\eqn{eq:PS_fact} where $[\rd p_{1}]$ is a one-parton phase space times
a Jacobian factor,
\beq
[\rd p_{1;m}] =
\Jac{}(p_r, \momt{};Q) \,\frac{\rd^d p_r}{(2\pi)^{d-1}}\delta_{+}(p_r^2)\,.
\label{eq:dp_1}
\eeq
With our definitions for the momentum mappings the Jacobian
factor depends on the number of hard final-state momenta.
In the case of the collinear mapping in \eqn{eq:PS_Cir}, the Jacobian is
\beq
\Jac{(ir)}(p_r,\ti{p}_{ir};Q) =
\frac{(1-\alpha_{ir})^{(m-1)(d-2)-1}y_{\wti{ir}Q}}
{\sqrt{(y_{r\wti{ir}}+y_{\wti{ir}Q}-y_{rQ})^2+4y_{r\wti{ir}}(1-y_{\wti{ir}Q})}}
\,\Theta(1-\alpha_{ir})\,,
\label{eq:Jac_Cir}
\eeq
and that for the soft mapping of \eqn{eq:PS_Sr} is
\beq
\Jac{(r)}(p_r;Q) =
\lambda_{r}^{(m-1)(d-2)-2}\,\Theta(\lambda_{r})
\,.
\label{eq:Jac_Sr}
\eeq
In \eqn{eq:Jac_Cir} $\alpha_{ir}$ is expressed in terms of the variable 
$\ti{p}_{ir}$,
\beq
\alpha_{ir}=
\frac{
\sqrt{(y_{r\wti{ir}}+y_{\wti{ir}Q}-y_{rQ})^2
     +4y_{r\wti{ir}}(1-y_{\wti{ir}Q})}
-(y_{r\wti{ir}}+y_{\wti{ir}Q}-y_{rQ})}{2(1-y_{\wti{ir}Q})}\,.
\label{eq:alpha_ir_new}
\eeq

This concludes the definition of the approximate cross section in
\eqn{eq:dsigRA1}. Note that our $\dsig{R,A}$ in general contains fewer
subtraction terms than the dipole scheme. Furthermore, we can decrease
the number of terms in \eqn{eq:bcA1}, because the symmetric treatment
of the collinear subtractions is convenient for bookkeeping, but not
essential in an actual computer code.

\section{Integral of the approximate cross section}
\label{sec:IntA1}

Next we evaluate the integral of the singly-unresolved approximate
cross section over the one-parton unresolved phase space. Let us begin
with integrating the collinear counterterm $\cC{ir}{}$.  The
transverse momentum $\kTt{i,r}$ as defined by \eqn{eq:kTtir} is
orthogonal to $\ti{p}_{ir}$, therefore, the spin correlations generally
present in \eqn{eq:Cir00} vanish after azimuthal integration
\cite{Catani:1996vz}.  Thus when evaluating the integral of the
subtraction term $\cC{ir}{}(\mom{})$ over the factorised one-parton
phase space $[\rd p_{1;m}^{(ir)}(p_r,\ti{p}_{ir};Q)]$, we may replace the
Altarelli--Parisi splitting functions $\hP^{}_{f_i f_r}$ by their
azimuthally averaged counterparts $P^{}_{f_i f_r}$. Then 
\beq
\int\,[\rd p_{1;m}^{(ir)}(p_r,\ti{p}_{ir};Q)] \cC{ir}{}(\mom{}) =
\aeps
\,\IcC{ir}{}(y_{\wti{ir}Q};m-1,\eps)\,\bT_{ir}^2\,\SME{m}{0}{\momt{(ir)}}\,,
\label{eq:IntCir00}
\eeq
where $y_{\wti{ir}Q} = 2 \ti{p}_{ir}\cdot Q$
and%
\footnote{We chose the dependence on $m$ in the argument of this
function as $m-1$ because it is due to dependence on $m$ in the
Jacobian factor in \eqn{eq:Jac_Cir}, where it appears as $m-1$ in the
exponent.}
\beq
\aeps \IcC{ir}{}(y_{\wti{ir}Q};m-1,\eps) = 
8\pi\as\mu^{2\eps}\int\,[\rd p_{1;m}^{(ir)}(p_r,\ti{p}_{ir};Q)]
\frac{1}{s_{ir}} P^{}_{f_i f_r}(\tzz{i}{r},\tzz{r}{i};\eps)
\frac{1}{\bT_{ir}^2}
\,.
\label{eq:ICir0}
\eeq
The evaluation of these integrals is discussed in
\app{app:single_integrals} and here we give only the final results. 

The function $\IcC{ir}{}(y_{\wti{ir}Q};n,\eps)$ depends on the momentum
of the  parent parton and the flavours of the daughter partons. The
three independent flavour combinations are  (we have
$\IcC{ir}{}(y_{\wti{ir}Q};n,\eps) = \IcC{ri}{}(y_{\wti{ir}Q};n,\eps)$
and $\IcC{\qb g}{}(y_{\wti{ir}Q};n,\eps) = \IcC{qg}{}(y_{\wti{ir}Q};n,\eps)$)
\beeq  
\IcC{qg}{}(x;n,\eps) \aand=   
x^{-2\eps}
\Big[\,2\Big({\rm I}^{(-1)}_n(x;\eps) - {\rm I}^{(0)}_n(x;\eps)\Big)  
+ (1-\eps){\rm I}^{(1)}_n(x;\eps)\Big],  
\label{eq:ICqg0}
\\  
\IcC{q\qb}{}(x;n,\eps) \aand=
\frac{\TR}{\CA}  
x^{-2\eps}
\left[\,{\rm I}^{(0)}_n(x;\eps) - \frac{2}{1-\eps}\Big({\rm I}^{(1)}_n(x;\eps) 
- {\rm I}^{(2)}_n(x;\eps)\Big)\right],  
\label{eq:ICqqb0}  
\eeeq  
and  
\beeq  
\IcC{gg}{}(x;n,\eps) \aand= 
2x^{-2\eps}
\Big[\,2\Big({\rm I}^{(-1)}_n(x;\eps) - {\rm I}^{(0)}_n(x;\eps)\Big)  
+ {\rm I}^{(1)}_n(x;\eps) - {\rm I}^{(2)}_n(x;\eps)\Big].  
\label{eq:ICgg0}  
\eeeq  
The $I^{(k)}_n(x;\eps)$ functions are given explicitly in
\Ref{DelDuca:2006rq}.

The expansion of \eqnss{eq:ICqg0}{eq:ICgg0} in powers of $\eps$ is
performed using the techniques of \cite{Moch:2001zr,Weinzierl:2002hv}
to obtain  
\beeq  
\IcC{qg}{}(x;n,\eps) \aand=
\left[\frac{1}{\eps^2}+\frac{3}{2\eps}-\frac{2}{\eps}\ln(x)+\Oe{0}\right],  
\label{eq:IcCqg0exp}  
\\  
\IcC{q\qb}{}(x;n,\eps) \aand=
\frac{\TR}{\CA}\left[-\frac{2}{3\eps}+\Oe{0}\right],  
\label{eq:IcCqqb0exp}  
\\  
\IcC{gg}{}(x;n,\eps) \aand=
\left[\frac{2}{\eps^2}+\frac{11}{3\eps}-\frac{4}{\eps}\ln(x)+\Oe{0}\right].  
\label{eq:IcCgg0exp}  
\eeeq  
The finite parts, not shown here, depend on $n$ and can be easily found
for any given $n$ using the program of Ref.~\cite{Weinzierl:2002hv}.
We quote those for $n=2,3$ and 4 in \app{app:single_integrals}.

Next consider the soft countertem. Defining
\beq
\aeps \IcS{ik}{}(y_{\ti{i}\ti{k}},y_{\ti{i}Q},y_{\ti{k}Q};m-1,\eps) = 
-8\pi\as\mu^{2\eps}\int [\rd p_{1;m}^{(r)}(p_r;Q)]
\,\frac12 \calS_{ik}(r)\,,
\label{eq:IcSik0}
\eeq
we obtain
\beeq
&&
\int [\rd p_{1;m}^{(r)}(p_r;Q)] \cS{r}{}(\mom{}) = 
\nn\\&&\qquad
\aeps
\,\sum_{i}\sum_{k\ne i}
\IcS{ik}{}(y_{\ti{i}\ti{k}},y_{\ti{i}Q},y_{\ti{k}Q};m-1,\eps)
\SME{m;(i,k)}{0}{\momt{(r)}}\,.
\label{eq:IntSr00}
\eeeq
Finally, integrating the soft-collinear subtraction, \eqn{eq:CirSr00} we
get
\beq
\int [\rd p_{1;m}^{(r)}(p_r;Q)] \cCS{ir}{r}{}(\mom{}) = 
\aeps \IcCS{}(m-1,\eps) \, \bT_i^2 \, \SME{m}{0}{\momt{(r)}}\,,
\label{eq:IntCirSr00}
\eeq
with
\beq
\aeps \IcCS{}(m-1,\eps) = 8\pi\as\mu^{2\eps}
\int [\rd p_{1;m}^{(r)}(p_r;Q)]
\,\frac{2}{s_{ir}}\frac{\tzz{i}{r}}{\tzz{r}{i}}\,.
\label{eq:IcCS0}
\eeq
The evaluation of the integrals in \eqns{eq:IcSik0}{eq:IcCS0} is again
discussed in \app{app:single_integrals} and here we give only the final
results.  

The soft functions
 $\IcS{ik}{}(y_{\ti{i}\ti{k}},y_{\ti{i}Q},y_{\ti{k}Q};n,\eps)$ are
expressed with the standard beta and hypergeometric functions
\cite{MathWorld}, 
\beeq
\IcS{ik}{}(y_{\ti{i}\ti{k}},y_{\ti{i}Q},y_{\ti{k}Q};n,\eps) \aand= 
-\frac{n(1-\eps)(1-2\eps)}{\eps^2} B(1-\eps,1-\eps)B(1-2\eps,n(1-\eps))  
\nn \\ &&\times  
\frac{y_{\ti{i}\ti{k}}}{y_{\ti{i}Q}y_{\ti{k}Q}} 
\,{}_2\!F_1
\left(1,1,1-\eps,1- \frac{y_{\ti{i}\ti{k}}}{y_{\ti{i}Q}y_{\ti{k}Q}}\right)\,,
\label{eq:ISik0}
\eeeq
while
\beq
\IcCS{}(n,\eps) =
\left[\frac{n(1-\eps)(1-2\eps)}{\eps^2} + 2\right]
B(1-\eps,1-\eps)B(1-2\eps,n(1-\eps))\,.  
\label{eq:ICS0}
\eeq
Using the expansion 
\beq
z\;{}_2\!F_1(1,1,1-\eps,1-z) =  
z^{-\eps}\,\left[1 + \eps^2 \Li_2(1-z) + \Oe{3}\right]  
\label{eq:2F1exp}
\eeq 
for the hypergeometric function, we find
\beeq
\IcS{ik}{}(y_{\ti{i}\ti{k}},y_{\ti{i}Q},y_{\ti{k}Q};n,\eps) \aand=
- \frac{1}{\eps^2}
- \frac{2}{\eps} \sum_{k=1}^n \frac{1}{k}
+ \frac{1}{\eps}\,\ln\frac{y_{\ti{i}\ti{k}}}{y_{\ti{i}Q}y_{\ti{k}Q}}
+ \Oe{0}\,,
\label{eq:ISik0exp}
\\[2mm]
\IcCS{}(n,\eps) \aand=
\frac{1}{\eps^2} + \frac{2}{\eps} \sum_{k=1}^n \frac{1}{k} + \Oe{0}\,.
\label{eq:ICS0exp}
\eeeq 
Notice that
\beq
\IcS{ik}{}(y_{\ti{i}\ti{k}},y_{\ti{i}Q},y_{\ti{k}Q};m,\eps)
+ \IcCS{}(m,\eps) =
\frac{1}{\eps} \,\ln\frac{y_{\ti{i}\ti{k}}}{y_{\ti{i}Q}y_{\ti{k}Q}}
+ \Oe{0}\,.
\label{eq:ISik0+CS0exp}
\eeq 
We quote the finite part of this expansion in
\app{app:single_integrals}.

We are now in a position to calculate $\int_1 \dsig{R,A}_{m+1}$. Let us
begin by recalling the form of the fully differential real cross
section $\dsig{R}_{m+1}$ given in \eqn{eq:dsigR}. Accordingly, the
approximate cross section times the jet function is
\beeq
&&
\dsig{R,A}_{m+1} J_{m} =
{\cal N}\sum_{\{m+1\}}\PS{m+1}{}(\mom{})\frac{1}{S_{\{m+1\}}}
\nn\\&&\qquad
\times
\sum_{r} \Bigg[\sum_{i\ne r} \frac{1}{2} \cC{ir}{}(\mom{}) J_{m}(\momt{(ir)})
\nn\\&&\qquad\qquad\quad
+ \left(\cS{r}{}(\mom{}) - \sum_{i\ne r} \cCS{ir}{r}{}(\mom{})\right) 
J_{m}(\momt{(r)})
\Bigg]\,.
\label{eq:dsigRA1_full}
\eeeq
In order to evaluate $\int_1 \dsig{R,A}_{m+1}$ we first use the phase
space factorization property of \eqn{eq:PS_fact},
then perform the integration to obtain
\beeq
&&
\int_1\dsig{R,A}_{m+1} J_{m} =
{\cal N}\sum_{\{m+1\}}\,\PS{m}{}(\momt{})\,\frac{1}{S_{\{m+1\}}}\,\aeps
\nn \\&&\qquad
\times
\sum_{r}\sum_{i\ne r}
\Bigg[\frac12 \IcC{ir}{}(m-1,\eps)
\,\bT_{ir}^2 \SME{m}{0}{\momt{}} J_{m}(\momt{})
\nn\\&&\qquad\qquad\qquad\quad
+\sum_{k\ne i,r} \IcS{ik}{}(m-1,\eps)\,\SME{m;(i,k)}{0}{\momt{}} J_{m}(\momt{})
\nn\\&&\qquad\qquad\qquad\quad
-\IcCS{}(m-1,\eps)\,\bT_i^2 \SME{m}{0}{\momt{}} J_{m}(\momt{})\Bigg]\,.
\label{eq:I1dsigRA_v2}
\eeeq
This result is not yet in the form of an $m$-parton contribution times a
factor. In order to rewrite \eqn{eq:I1dsigRA_v2} in such a form we
still need to perform the counting of symmetry factors for going from
$m$ partons to $m+1$ partons, which is very similar to the counting in
\Ref{Catani:1996vz}. We give the details of the calculation in
\app{app:symfactors}. Inserting equation \eqn{eq:m+1_sum_to_m_sum_fin}
into \eqn{eq:I1dsigRA_v2}, we obtain
\beeq
&&
\int_1\dsiga{RR}{1}_{m+1} J_{m} =
{\cal N}\sum_{\{m\}}\PS{m}{}(\momt{})\frac{1}{S_{\{m\}}} \aeps
J_{m}(\momt{})
\nn \\&&\quad\times
\sum_{i}
\Bigg[
  \IcC{i}{}(m-1,\eps) \,\bT_i^2\, \SME{m}{0}{\momt{}}
+ \sum_{k\ne i}
\IcS{ik}{}(m-1,\eps)
\SME{m;(i,k)}{0}{\momt{}}
\Bigg]
\,,\qquad~
\label{eq:I1dsigRA_v4}
\eeeq
where we have introduced the flavour-dependent functions
\beq
\IcC{q}{} = \IcC{qg}{} -  \IcCS{}\,,\qquad 
\IcC{g}{} = \frac12 \IcC{gg}{} + \Nf \IcC{q\qb}{} -  \IcCS{}\,.
\label{eq:IcCi_def}
\eeq

We write the final result, dropping the $m$-parton jet function that
appears on both sides of \eqn{eq:I1dsigRA_v4}, as follows
\beq
\int_1\dsig{R,A}_{m+1} = \dsig{B}_{m} \otimes \bI(m-1,\eps)\,,
\label{eq:I1dsigRA_fin2}
\eeq 
where $\dsig{B}_{m}$ is the Born-level cross section for the emission
of $m$ partons and 
\beeq
&&
\bI(\mom{};m-1,\eps) = 
\aeps
\nn\\&&\quad
\times
  \sum_{i} \left[ \IcC{i}{}(y_{iQ};m-1,\eps) \,\bT_i^2
+ \sum_{k\ne i} \IcS{ik}{}(y_{ik},y_{iQ},y_{kQ};m-1,\eps)\,\bT_i \bT_k \right]
\label{eq:bI_def}
\eeeq
is an operator acting on the colour space of $m$ partons that depends
on the colour charges and momenta of the $m$ partons in $\M{m}{}$
(different from the insertion operator of \Ref{Catani:1996vz}).  The
notation on the right hand side of \eqn{eq:I1dsigRA_fin2} means that
one has to write down the expression for $\dsig{B}_{m}$ and then
replace the Born level squared matrix element
\beq
\M{m}{} = \la {\cal M}_{m}^{(0)}|{\cal M}_{m}^{(0)} \ra\,,
\eeq
with
\beq
\la {\cal M}_{m}^{(0)}|\bI(m-1,\eps)|{\cal M}_{m}^{(0)} \ra\,.
\label{eq:MmIMm}
\eeq
The parameter $m-1$ in the argument of the insertion operator matches the
arguments of the collinear and soft functions, not the number of coloured
legs in the matrix element (in this case it is one less).
Using colour conservation $(\bT_i^2 = -\sum_{k\ne i}\bT_i \bT_k)$ to
combine the $\IcC{i}{}$ and $\IcS{ik}{}$ contributions,
\eqnss{eq:IcCqg0exp}{eq:IcCgg0exp} and \eqn{eq:ISik0+CS0exp},
it is straightforward to check that our insertion operator differs from
that defined in (7.26) of \Ref{Catani:1996vz} only in finite terms,
\beq
\bI(\mom{};m-1,\eps) = \bI(\mom{};\eps) + \Oe{0}\,,
\label{eq:bI_poles}
\eeq
where
\beq
\bI(\mom{};\eps) = \aeps
\sum_i\left( \bT_i^2 \frac{1}{\eps^2} + \gamma_i \frac{1}{\eps}
+ \sum_{k\ne i} \bT_i \bT_k \frac{1}{\eps} \ln y_{ik}\right)\,
\label{eq:bI}
\eeq
with the usual flavour constants
\beq
\gamma_q = \frac32 \CF\,,\qquad \gamma_g = \frac{\beta_0}{2}\,.
\label{eq:gammadef}
\eeq
It follows that $\int_1 \dsig{R,A}_{m+1}$, as defined here, correctly
cancels all $\eps$-poles of the virtual cross section $\dsig{V}_{m}$.
As a result, the integrand of the $m$-parton contribution,
\beq
\dsig{NLO}_m =
\left[\dsig{V}_m + \dsig{B}_{m} \otimes \bI(m-1,\eps)\right]_{\eps = 0} J_m
\:,
\eeq
is finite and integrable in four dimensions (the potential kinematical
singularities are screened by the jet function $J_m$). This finite
integrand is given in \app{app:single_integrals}.

\section{Checks}
\label{sec:checks}

The cancellation of the singularities is a strong check on the
correctness of the proposed scheme. We have performed such checks by
approaching soft or collinear regions of the phase space from a
randomly chosen point and computing the ratio of the $(m+1)$ parton
squared matrix element and the subtraction terms. This ratio always
approaches one. Further checks can be performed by comparing
predictions for distributions to the predictions of other
well-established computer codes for computing QCD jet cross sections at
the NLO accuracy. Currently, our scheme is worked out only for
colourless particles in the initial state, therefore, we decided to
compare the predictions for the three-jet event-shape distributions
thurst ($T$) and $C$-parameter in electron-positron annihilation, when
the jet function is a functional
\beq
J_n(p_1,\ldots,p_n;O) = \delta(O-O_3(p_1,\ldots,p_n))\:,
\eeq
with $O_3(p_1,\ldots,p_n)$ being the value of either $\tau \equiv 1-T$
or $C$ for a given event $(p_1,\ldots,p_n)$.  

Once the phase space integrations in \eqn{eq:sigmaNLOfin} are carried
out, the NLO differential cross section for the three-jet observable
$O$ at a fixed scale $Q$ takes the general form
\beq
\Sigma(O)\equiv O \frac{1}{\sigma_0}\frac{\rd \sigma}{\rd O}(O)
= \frac{\as(Q)}{2\pi} B_{O}(O) 
+ \left(\frac{\as(Q)}{2\pi}\right)^2 C_{O}(O)\:.
\label{nloxsec}
\eeq
We computed the $B_{O}(O)$ Born-level predictions as well as the
$C_{O}(O)$ correction functions and found complete agreement with the
corresponding tables of the benchmark calculation of Kunszt and Nason
\cite{Kunszt:1989km}.

\section{Conclusions}

We have defined a new subtraction scheme for computing NLO corrections to
QCD jet cross sections. For NLO computations the new scheme does not
provide any particular advantage over the already existing methods and
gives identical predictions. The need for the new scheme was motivated
by studies in trying to extend the existing NLO subtraction schemes to
computations at the NNLO accuracy.

The new scheme is completely general in the sense that any number of
massless final state coloured or colourless particles are allowed. It
is worked out for processes without coloured partons in the initial
state. The extension to deep-inelastic scattering and hadron collisions
does not pose conceptual difficulties, but left for later work.

\section*{Acknowledgments}
We are grateful to Prof. Lovas for introducing us into theoretical
particle physics and for the hospitality of the CERN Theory Division
where this work was completed.
This research was supported by the Hungarian Scientific Research Fund
grant OTKA K-60432.  

\appendix

\section{Integrals over the factorised single-particle phase space}
\label{app:single_integrals}

In this appendix we compute the collinear and soft functions
$\IcC{ir}{}$, $\IcS{r}{}$ and $\IcCS{}$ and present their expansions
relevant to three-, four- and five-jet production.

\subsection{Collinear integrals}

We recall the definition of the collinear functions, \eqn{eq:ICir0}, from
which one trivially gets
\beq
\IcC{ir}{}(y_{\wti{ir}Q};m-1,\eps) = 
\frac{(4\pi)^2}{S_\eps}\,(Q^2)^{-1+\eps}
\int\,[\rd p_{1;m}^{(ir)}(p_r,\ti{p}_{ir};Q)]
\frac{1}{y_{ir}} P^{}_{f_i f_r}(\tzz{i}{r},\tzz{r}{i};\eps)
\frac{1}{\bT_{ir}^2}
\,,
\label{eq:ICir0def}
\eeq
where the azimuthally averaged splitting kernels are
\beeq
P_{g_ig_r}(z_r) \aand=
2\CA\left[\frac{1-z_r}{z_r} + \frac{z_r}{1-z_r} + (1-z_r) z_r \right]\,,
\label{eq:P0gg}
\\
P_{\qb_i q_r}(z_r;\eps) \aand=
\TR\left[
1 - \frac{2}{1-\eps} (1-z_r) z_r
\right]\,,
\label{eq:P0qq}
\\
P_{q_ig_r}(z_r;\eps) \aand=
\CF\left[\frac{1+(1-z_r)^2}{z_r}-\eps z_r\right]
\,.
\label{eq:P0qg}
\eeeq
We parametrise the factorised one-particle phase space in terms of the
variables $\alpha_{ir}$, $y_{ir}$ and momentum fraction $z_r \equiv
\tzz{r}{i}$, the latter being defined in \eqn{eq:zt2}. We find
\beeq
[\rd p_{1;m}^{(ir)}(p_r,\ti{p}_{ir};Q)] \aand= (1-\alpha_{ir})^{[(m-1)(d-2)-1]}
\,y_{\wti{ir}Q}
\,\left(Q^2\right)^{1-\eps}\frac{S_\eps}{(4\pi)^2}
\nn\\ \aand \times
\,\delta\left(
y_{ir} - \alpha_{ir}\,(\alpha_{ir} + y_{\wti{ir}Q} - \alpha_{ir} y_{\wti{ir}Q})
\right)
\,\rd \alpha_{ir}\,\rd y_{ir}\,\rd z_r
\nn\\ \aand \times
\,(z_+-z_-)^{-1+2\eps}
\,[y_{ir}\,(z_+-z_r)\,(z_r-z_-)]^{-\eps}
\nn\\ \aand \times
\,\Theta(1-\alpha_{ir})\,\Theta(\alpha_{ir})
\,\Theta(z_+-z_r)\,\Theta(z_r-z_-)\,,
\label{eq:PS_single_coll_fact2}
\eeeq
The limits of the $z_r$-integral are
\beeq
z^{(+)}(\alpha_{ir}, y_{ir}, y_{\wti{ir}Q}) \aand= \frac{y_{ir}}
{\alpha_{ir}\,(2\alpha_{ir} + y_{\wti{ir}Q} - \alpha_{ir} y_{\wti{ir}Q})}
\,,
\nn\\
z^{(-)}(\alpha_{ir}, y_{ir}, y_{\wti{ir}Q}) \aand= \frac{y_{ir}}
{(\alpha_{ir} + y_{\wti{ir}Q} - \alpha_{ir} y_{\wti{ir}Q})
 (2\alpha_{ir} + y_{\wti{ir}Q} - \alpha_{ir} y_{\wti{ir}Q})}
\,,
\label{eq:zrpzrm}
\eeeq
with $z^{(+)} + z^{(-)} = 1$.  We now insert
\eqnss{eq:P0gg}{eq:PS_single_coll_fact2} into \eqn{eq:ICir0def}
and obtain the results presented in \eqnss{eq:ICqg0}{eq:ICgg0}, with 
\beeq
x^{-2\eps}\,{\rm I}^{(k)}_n(x;\eps) \aand= 
\int_0^1\!\rd \alpha\, \int_0^1\!\rd y\,
\int_{z_-(\alpha,y,x)}^{z_+(\alpha,y,x)}\!\rd z
\,\delta\left( y - \alpha\,(\alpha + x - \alpha x) \right)
\nn\\ && \qquad\times
\,(1-\alpha)^{[2n(1-\eps)-1]}
\,[z_+(\alpha,y,x)-z_-(\alpha,y,x)]^{-1+2\eps}
\nn\\ && \qquad\times
\,[y\,(z_+(y,\alpha,x)-z)\,(z-z_-(y,\alpha,x))]^{-\eps}
\;\frac{x}{y} z^k
\,.
\label{eq:Inm}
\eeeq
These integrals are invariant under the $z \leftrightarrow 1-z$
transformation, therefore, not all are independent; those with positive
and odd $n$ can be expressed with the others. For instance,
${\rm I}^{(1)}_n(x;\eps) = \frac12\,{\rm I}^{(0)}_n(x;\eps)$.
The integral over $y$ is trivial by making use of the $\delta$ function.
After a complicated sequence of intergal transformations, the other two
integrals can be transformed into known integrals. The results can be
found in \Ref{DelDuca:2006rq}.

In an actual computation we need the expansion of the collinear functions
in $\eps$ to $\Oe{0}$. The pole terms are independent of $m$ and are
given in \eqnss{eq:IcCqg0exp}  {eq:IcCgg0exp}. Here we give the
$\Oe{0}$ terms, denoted by $\finite \IcC{ir}{}$.
For $n = m-1 = 1$ we need the functions only at $x=1$, where
\beq
\finite \IcC{qg}{}(1;1) = 
7 - \frac{\pi^2}{2}
\,,
\qquad 
\finite \IcC{q\qb}{}(1;1) = 
\frac{2\TR}{3\CA}\, \frac{3\ln 2 - 11}{3}
\,.
\eeq
For $n = m-1 \geq 2$ the results for arbitrary $n$ are somewhat
cumbersome combinations of elementary, ${}_2\!F_1$ and ${}_3\!F_2$
functions. Equivalent simpler expressions can be given in the following
way:
\beeq
\finite \IcC{qg}{}(x;n) \aand= 
  c_n^{qg}
+ 2 \left(\ln^2 x + \Phi(1 - x, 2, 2n-1) + \Li_2(1-x) - \frac{\pi^2}{4}\right)
\nn\\&&
+ \Big(d_{n,0} - 3\Big) \ln x 
- \sum_{i=1}^{n-1} 
  \frac{d_{n,i}}{i}\,{}_2\!F_1(i, 1, 1 + i, 1 - x)
\nn\\&&
+ \sum_{i=0}^{n-1} 
  \frac{d_{n,i}}{2n-1-i}\,{}_2\!F_1(2 n - 1 - i, 1, 2 n - i, 1 - x)
\,,
\\[2mm]
\finite \IcC{q\qb}{}(x;n) \aand= 
\frac{2\TR}{3\CA} \Bigg(
c_n^{qq} +  \ln x
- \frac{1}{2 n-1}\,{}_2\!F_1(2 n-1, 1, 2 n, 1 - x)
\nn\\&&\qquad\qquad
+ \frac{2 n - 1}{4 n}\,\frac{x}{2}
\,{}_2\!F_1\left(1, 2 n, 2 n + 1, \frac{2 - x}2\right)
\Bigg)
\,.
\eeeq
The finite parts of the collinear functions for the gluon are not
independent from the other two:
\beq
\finite \IcC{gg}{}(x;n) = 
2 \finite \IcC{qg}{}(x;n)
- \frac{\CA}{\TR} \finite \IcC{q\qb}{}(x;n)
- \frac23
\,.
\eeq
The constants $c_n^{ir}$ and $d_{n,i}$ can be found in \tab{tab:constants}.
The function $\Phi(z,s,a)$ is the Lerch transcendent \cite{MathWorld},
defined by the series
\beq
\Phi(z,s,a) = \sum_{k=0}^\infty \frac{z^k}{(a+k)^s}
\,,
\eeq
for which numerical codes for the evaluation exist \cite{lerchphi}.
While numerical codes for the hypergeometric functions also exist, it is
actually faster to expand the ${}_2\!F_1$ functions using their
series expansion because for integer arguments the representation is a
sum of polinoms and logarithms.

\begin{table}
\caption{Constants used in the finite part of the expansion of the
collinear functions $C_{ir}$}
\label{tab:constants}
\begin{center}
\begin{tabular}{cccccccc}
\hline
\hline
$m$ & $n$ & $c_n^{qg}$ & $c_n^{q\qb}$ & 
$d_{n,0}$ & $d_{n,1}$ & $d_{n,2}$ & $d_{n,3}$ \\
\\
\hline
\\[-2mm]
3&2&$\frac{19}{4}$   & $-\frac{11}{3}$  &$-\frac32$& 1 &   &   \\
\\[-2mm]
4&3&$\frac{89}{24}$  & $-\frac{17}{4}$  &$-\frac83$&$\frac32$&$\frac13$&   \\
\\[-2mm]
5&4&$\frac{959}{360}$& $-\frac{277}{60}$&$-\frac{17}{5}$&$\frac53$&$\frac35$
&$\frac16$\\
\\[-2mm]
\hline
\hline
\end{tabular} 
\end{center} 
\end{table}

\subsection{Soft integrals}

We recall the definition of the soft functions, \eqns{eq:IcSik0}{eq:IcCS0},
from which one trivially gets
\beq
\IcS{ik}{}(y_{\ti{i}\ti{k}},y_{\ti{i}Q},y_{\ti{k}Q};m-1,\eps) = 
-\frac{(4\pi)^2}{S_\eps}\,(Q^2)^{-1+\eps}
\int [\rd p_{1;m}^{(r)}(p_r;Q)]
\,\frac{y_{ik}}{y_{ir} y_{kr}}
\label{eq:IcSik0def}
\eeq
and
\beq
\IcCS{}(m-1,\eps) = 
\frac{(4\pi)^2}{S_\eps}\,(Q^2)^{-1+\eps}
\int [\rd p_{1;m}^{(r)}(p_r;Q)]
\,\frac{2}{y_{ir}}\frac{\tzz{i}{r}}{\tzz{r}{i}}\,.
\label{eq:IcCS0def}
\eeq
Parametrizing the phase space with energy and angles, these integrals
can be computed as done in Appendix B of \Ref{Catani:1996vz} (see also
\cite{Beenakker:1988bq}), leading to \eqns{eq:ISik0}{eq:ICS0}. Although, 
in \eqn{eq:IcCi_def} we have combined the $\IcCS{}$ functions with the
collinear ones because these are multiplied with the same colour factor,
nevertheless one can always use colour conservation
$(\bT_i^2 = -\sum_{k\ne i}\bT_i \bT_k)$ to combine the $\IcCS{}$ and
$\IcS{ik}{}$ contributions. Therefore, in presenting the $\Oe{}$ terms 
in the $\eps$-expansion, we write only the finite term of their sum,
which is simpler than the individual contributions,
\beq
\finite [\IcS{ik}{}(y_{\ti{i}\ti{k}},y_{\ti{i}Q},y_{\ti{k}Q};n)
+ \IcCS{}(n)] =
  \frac{2}{n}
+ \Li_2\left(1 - \frac{y_{\ti{i}Q} y_{\ti{k}Q}}{y_{\ti{i}\ti{k}}}\right)
+ 2\ln \left(\frac{y_{\ti{i}\ti{k}}}{y_{\ti{i}Q} y_{\ti{k}Q}}\right)
\sum_{k=1}^n \frac1k
\,.
\eeq

In order to spell out the finite part of the $m$-parton contribution,
$ \dsig{NLO}_m$, we define the finite part of the one-loop amplitude as
\beq
\ket{m}{(1)}{(\mom{})} =
- \frac12 \bI(\mom{};\eps) \ket{m}{(0)}{(\mom{})}
+ \finite \ket{m}{(1)}{(\mom{})}
\,.
\eeq
Then
\beeq
&&
\left[\dsig{V}_m + \dsig{B}_{m} \otimes \bI(m-1,\eps)\right]_{\eps = 0} =
{\cal N}\sum_{\{m\}}\PS{m}{}
\,\frac{1}{S_{\{m\}}}
\Bigg\{
2 \Real \bra{m}{(0)}{(\mom{})}\finite \ket{m}{(1)}{(\mom{})}
\nn\\&&\qquad
+ \frac{\as}{2\pi}
  \sum_{i} \Bigg[ \sum_{k\ne i}
  \finite\left[\IcS{ik}{}(y_{ik},y_{iQ},y_{kQ};m-1) + \IcCS{}(m-1)\right]
\,\SME{m;(i,k)}{0}{\mom{}}
\nn\\&&\qquad\qquad\qquad
+ \finite\left[\IcC{i}{}(y_{iQ};m-1)  + \IcCS{}(m-1)\right]\,\bT_i^2
\,\SME{m}{0}{\mom{}}
\Bigg]
\Bigg\}
\,.
\qquad~
\eeeq

\section{Calculation of symmetry factors}
\label{app:symfactors}

Consider an $m$-parton configuration with $m_f$ quarks of flavour
$f$, $\bar{m}_f$ antiquarks of flavour $f$ and $m_g$ gluons. From this
configuration we can obtain an $m+1$ parton configuration by changing
\beq
\mbox{(i)}\,\,\, m_g \to m_g + 1
\qquad \mbox{or} \qquad
\mbox{(ii)}\,\,\, m_f \to m_f + 1, \, \bar{m}_f \to \bar{m}_f + 1, \, m_g \to m_g - 1\,.
\label{eq:m+1tom+2}
\eeq
The ratios of symmetry factors corresponding to the two cases are
\beeq
\frac{S_{\{m\}}^{(\rm{i})}}{S_{\{m+1\}}} \aand=
\frac{\ldots m_g!}{\ldots (m_g + 1)!}
= \frac{1}{m_g + 1}\,,
\label{eq:Sratios}
\\
\frac{S_{\{m\}}^{(\rm{ii})}}{S_{\{m+1\}}} \aand= 
\frac{\ldots m_f!\bar{m}_f!m_g!}{\ldots (m_f + 1)!(\bar{m}_f + 1)!(m_g - 1)!}
= \frac{m_g}{(m_f + 1)(\bar{m}_f + 1)}\,.
\nn
\eeeq
We then have
\beeq
\sum_{\{m+1\}}\frac{1}{S_{\{m+1\}}}\sum_i\sum_{r\ne i}\ldots \aand=
\sum_{\{m\}}\!\strut^{(\rm{i})}\frac{1}{S_{\{m\}}}\frac{1}{m_g+1}
\Bigg( \sum_{i=q_f}\sum_{r=g}\ldots + \sum_{i=g}\sum_{r=q_f}\ldots
\label{eq:m+1_sum_to_m_sum}
\\&&
\nn
      +\sum_{i=\qb_f}\sum_{r=g}\ldots + \sum_{i=g}\sum_{r=\qb_f}\ldots
      +\sum_{i=g}\sum_{r=g}\ldots\Bigg)
\nn\\ &&
+\sum_{\{m\}}\!\strut^{(\rm{ii})}\frac{1}{S_{\{m\}}}
\frac{m_g}{(m_f+1)(\bar{m}_f+1)}
\Bigg( \sum_{i=q_f}\sum_{r=\qb_f}\ldots + \sum_{i=\qb_f}\sum_{r=q_f}\ldots
\Bigg).
\qquad~
\eeeq
Also
\beeq
\sum_{i=q_f}\sum_{r=g}\ldots \aand= (m_g+1)\sum_{\wti{ir}=q_f}\ldots \,, \qquad
\sum_{i=g}\sum_{r=q_f}\ldots = (m_g+1)\sum_{\wti{ir}=q_f}\ldots \,,
\nn\\
\sum_{i=\qb_f}\sum_{r=g}\ldots \aand= (m_g+1)\sum_{\wti{ir}=\qb_f}\ldots \,, \qquad
\sum_{i=g}\sum_{r=\qb_f}\ldots = (m_g+1)\sum_{\wti{ir}=\qb_f}\ldots \,,
\nn\\
\sum_{i=g}\sum_{r=g}\ldots \aand= (m_g+1)\sum_{\wti{ir}=g}\ldots \,,
\nn\\
\sum_{i=q_f}\sum_{r=\qb_f}\ldots \aand= \frac{(m_f+1)(\bar{m}_f+1)}{m_g}
\sum_{\wti{ir}=g}\ldots \,, \qquad
\nn\\
\sum_{i=\qb_f}\sum_{r=q_f}\ldots \aand= \frac{(m_f+1)(\bar{m}_f+1)}{m_g}
\sum_{\wti{ir}=g}\ldots \,,
\label{eq:irtoir}
\eeeq
thus we find
\beeq
\sum_{\{m+1\}}\frac{1}{S_{\{m+1\}}}\sum_i\sum_{r\ne i}\ldots \aand=
\sum_{\{m\}}\!\strut^{(\rm{i})}\frac{1}{S_{\{m\}}}
\Bigg( \sum_{\wti{ir}=q_f,r=g}\ldots + \sum_{\wti{ir}=g,r=q_f}\ldots
\nn \\&&
      +\sum_{\wti{ir}=\qb_f,r=g}\ldots + \sum_{\wti{ir}=g,r=\qb_f}\ldots
      +\sum_{\wti{ir}=g,r=g}\ldots\Bigg)
\nn\\ &&
+\sum_{\{m\}}\!\strut^{(\rm{ii})}\frac{1}{S_{\{m\}}}
\Bigg( \sum_{\wti{ir}=g,r=\qb_f}\ldots + \sum_{\wti{ir}=g,r=q_f}\ldots \Bigg)\,.
\label{eq:m+1_sum_to_m_sum_fin}
\eeeq
The soft contribution to each sum in \eqn{eq:m+1_sum_to_m_sum} is
nonvanishing only if $r$ is a gluon, so we have indicated the flavour
of $r$ in each summation in \eqn{eq:m+1_sum_to_m_sum_fin}.

\section{Volume of the phase space in $d$ dimensions}
\label{app:PS}

In this appendix, we present a simple derivation of the formula, obtained
in \Ref{Gehrmann-DeRidder:2003bm}, for the volume of the phase space of
$m$ massless particles in $d$ dimensions, which is a side product of the
phase-space factorization presented in \eqns{eq:dp_1}{eq:Jac_Sr}.

According to \eqns{eq:PS_fact}{eq:dp_1}
\beq
\int \PS{m+1}{(Q)} = \int \PS{m}{(Q)}\,I(Q;m,\eps)
\,,
\eeq
where, using \eqns{eq:lambdar}{eq:Jac_Sr},
\beq
I(Q;m,\eps) = \int 
\!(1-y_{rQ})^{\frac12[(m-1)(d-2)-2]}
\,\frac{\rd^d p_r}{(2\pi)^{d-1}}\delta_{+}(p_r^2)
\,.
\eeq
Working in the c.m.~frame, we parametrise the one-particle phase-space
measure with the energy and angles. The integrand, $(1-y_{rQ}) = (1-2 E/Q)$,
depends only on the energy, therefore,
\beq
I(Q;m,\eps) =
\frac{\Omega_{d-1}}{(2\pi)^{d-1}}
\int_0^{Q/2} \frac12 E^{d-3}\,\rd E
\,\left(1-2\frac{E}{Q}\right)^{\frac12[(m-1)(d-2)-2]}
\,,
\label{eq:IQ}
\eeq
where 
\beq
\Omega_d = \frac{2 \pi^{d/2}}{\Gamma(d/2)}
\eeq
is the volume of the $d$-dimensional hypersurface. Introducing the new
variable $x = 2 E/Q$, the integral in \eqn{eq:IQ} is readily obtained,
\beq
I(Q;m,\eps) =
(Q^2)^{\frac{d-2}{2}}
\,\frac{2^{3-2d} \pi^{(1-d)/2}}{\Gamma((d-1)/2)}
\,B\Big(d-2,[(m-1)(d-2)]/2\Big)
\,.
\eeq
We can get rid of the $\sqrt{\pi}$ factors by using the identity,
\beq
\sqrt{\pi} = 2^{d-3}
\frac{\Gamma\left(\frac{d-2}{2}\right)\Gamma\left(\frac{d-1}{2}\right)}
     {\Gamma(d-2)}
\,.
\eeq
Thus, we derived the following recursion relation:
\beq
\int \PS{m+1}{(Q)} =
2^{-d} \pi^{d/2} (Q^2)^{\frac{d-2}{2}}
\frac{\Gamma\left(\frac{(m-1)(d-2)}{2}\right)\Gamma\left(\frac{d-2}{2}\right)}
     {\Gamma\left(\frac{(m+1)(d-2)}{2}\right)}
\int \PS{m}{(Q)}
\,.
\label{eq:PSrecursion}
\eeq
Starting from the known expression for the two-particle phase space
\beq
\int \PS{2}{(Q)} =
2^{1-d} \pi^{1-d/2} (Q^2)^{\frac{d-4}{2}}
\frac{\Gamma\left(\frac{d-2}{2}\right)}{\Gamma(d-2)}
\,,
\eeq
and using the recursion relation in \eqn{eq:PSrecursion}, it is easy to
obtain the result quoted in \Ref{Gehrmann-DeRidder:2003bm} for $d = 4-2\eps$.


\begin{thebibliography}{99}

\bibitem{Giele:1991vf}
W.~T.~Giele and E.~W.~N.~Glover,
{\em Higher order corrections to jet cross-sections in e+ e-
annihilation},
Phys.\ Rev.\ D {\bf 46} (1992) 1980.

\bibitem{Giele:1993dj}
W.~T.~Giele, E.~W.~N.~Glover and D.~A.~Kosower,
{\em Higher order corrections to jet cross-sections in hadron colliders},
Nucl.\ Phys.\ B {\bf 403} (1993) 633
[arXiv:hep-ph/9302225].

\bibitem{Frixione:1995ms}
S.~Frixione, Z.~Kunszt and A.~Signer,
{\em Three-jet cross sections to next-to-leading order},
Nucl.\ Phys.\ B {\bf 467} (1996) 399
[arXiv:hep-ph/9512328].

\bibitem{Nagy:1996bz}
Z.~Nagy and Z.~Tr\'ocs\'anyi,
{\em Calculation of QCD jet cross sections at next-to-leading order},
Nucl.\ Phys.\ B {\bf 486} (1997) 189
[arXiv:hep-ph/9610498].

\bibitem{Frixione:1997np}
S.~Frixione,
{\em A general approach to jet cross sections in QCD},
Nucl.\ Phys.\ B {\bf 507} (1997) 295
[arXiv:hep-ph/9706545].

\bibitem{Catani:1996vz}
S.~Catani and M.~H.~Seymour,
{\em A general algorithm for calculating jet cross sections in NLO QCD},
Nucl.\ Phys.\ B {\bf 485} (1997) 291
[Erratum-ibid.\ B {\bf 510} (1997) 291]
[hep-ph/9605323].

\bibitem{Campbell:2000bg}
J.~M.~Campbell and R.~K.~Ellis,
{\em Radiative corrections to Z b anti-b production},
Phys.\ Rev.\ D {\bf 62}, 114012 (2000)
[arXiv:hep-ph/0006304].

\bibitem{Campbell:2002tg}
J.~Campbell and R.~K.~Ellis,
{\em Next-to-leading order corrections to W + 2jet and Z + 2jet
production  at hadron colliders},
Phys.\ Rev.\ D {\bf 65}, 113007 (2002)
[arXiv:hep-ph/0202176].

\bibitem{Campbell:2004ch}
J.~Campbell, R.~K.~Ellis and F.~Tramontano,
{\em Single top production and decay at next-to-leading order},
Phys.\ Rev.\ D {\bf 70}, 094012 (2004)
[arXiv:hep-ph/0408158].

\bibitem{Nagy:2003tz}
Z.~Nagy,
{\em Next-to-leading order calculation of three-jet observables in hadron
hadron collision},
Phys.\ Rev.\ D {\bf 68}, 094002 (2003)
[arXiv:hep-ph/0307268].

\bibitem{Nagy:2001xb}
Z.~Nagy and Z.~Tr\'ocs\'anyi,
{\em Multi-jet cross sections in deep inelastic scattering at
next-to-leading order},
Phys.\ Rev.\ Lett.\  {\bf 87}, 082001 (2001)
[arXiv:hep-ph/0104315].

\bibitem{Nagy:1998bb}
Z.~Nagy and Z.~Tr\'ocs\'anyi,
{\em Next-to-leading order calculation of four-jet observables in
electron positron annihilation},
Phys.\ Rev.\ D {\bf 59}, 014020 (1999)
[Erratum-ibid.\ D {\bf 62}, 099902 (2000)]
[arXiv:hep-ph/9806317].

\bibitem{Somogyi:2006_2}
G.~Somogyi, Z.~Tr\'ocs\'anyi and V.~Del Duca,
{\em A subtraction scheme for computing QCD jet cross sections at NNLO:
regularization of doubly-real emissions},
[arXiv:hep-ph/0609042].

\bibitem{Somogyi:2006_3}
G.~Somogyi and Z.~Tr\'ocs\'anyi,
{\em A subtraction scheme for computing QCD jet cross sections at NNLO:
regularization of real-virtual emissions},
[arXiv:hep-ph/0609043].

\bibitem{Altarelli:1977zs}
G.~Altarelli and G.~Parisi,
{\em Asymptotic Freedom In Parton Language},
Nucl.\ Phys.\ B {\bf 126}, 298 (1977).

\bibitem{Mangano:1990by}
M.~L.~Mangano and S.~J.~Parke,
{\em Multiparton Amplitudes In Gauge Theories},
Phys.\ Rept.\  {\bf 200}, 301 (1991).

\bibitem{Somogyi:2005xz}
G.~Somogyi, Z.~Tr\'ocs\'anyi and V.~Del Duca,
{\em Matching of singly- and doubly-unresolved limits of tree-level QCD
squared matrix elements},
JHEP {\bf 0506}, 024 (2005)
[arXiv:hep-ph/0502226].

\bibitem{DelDuca:2006rq}
V.~Del Duca, G.~Somogyi and Z.~Trocsanyi,
{\em Progress on NNLO subtraction},
Nucl.\ Phys.\ Proc.\ Suppl.\  {\bf 157}, 37 (2006)
[arXiv:hep-ph/0602068].

\bibitem{Moch:2001zr}
S.~Moch, P.~Uwer and S.~Weinzierl,
{\em Nested sums, expansion of transcendental functions and multi-scale
multi-loop integrals}
J.\ Math.\ Phys.\  {\bf 43}, 3363 (2002)
[arXiv:hep-ph/0110083].

\bibitem{Weinzierl:2002hv}
S.~Weinzierl,
{\em Symbolic Expansion of Transcendental Functions},
Comput.\ Phys.\ Commun.\  {\bf 145}, 357 (2002)
[arXiv:math-ph/0201011].

\bibitem{MathWorld}
http://mathworld.wolfram.com/

\bibitem{Kunszt:1989km}
Z.~Kunszt, P.~Nason, G.~Marchesini and B.~R.~Webber,
{\em QCD At Lep},
ETH-PT-89-39,
in {\it Proceedings of the 1989 LEP Physics Workshop,
Geneva, Swizterland, Feb 20, 1989}

\bibitem{lerchphi}
See for instance, Sergej V. Aksenov and Ulrich D. Jentschura,
{\sc lerchphi.c},
http://www.mpi-hd.mpg.de/personalhomes/ulj/jentschura/Source/lerchphi.c

\bibitem{Beenakker:1988bq}
W.~Beenakker, H.~Kuijf, W.~L.~van Neerven and J.~Smith,
{\em QCD Corrections To Heavy Quark Production In P Anti-P Collisions},
Phys.\ Rev.\ D {\bf 40}, 54 (1989).

\bibitem{Gehrmann-DeRidder:2003bm}
A.~Gehrmann-De Ridder, T.~Gehrmann and G.~Heinrich,
{\em Four-particle phase space integrals in massless QCD},
Nucl.\ Phys.\ B {\bf 682}, 265 (2004)
[arXiv:hep-ph/0311276].

\end{thebibliography}
\end{document}